\documentclass[10pt,a4papper]{article}
\usepackage[utf8x]{inputenc}
\usepackage[T2A]{fontenc}
\usepackage{indentfirst}
\usepackage{amsmath}
\usepackage{amssymb}
\usepackage[english]{babel}
\usepackage[]{graphicx}
\usepackage{geometry}
\usepackage{float}

\begin{document}

\title{\textbf{The decays $\tau\rightarrow (K, K(1460))\nu_{\tau}$ and the value of the weak decay coefficients  ${F}_{K}$ and ${{F}_{{K}^{'}}}$ in the extended NJL model}}
\author{M. K. Volkov$^{1}$\footnote{volkov@theor.jinr.ru}, K. Nurlan$^{1,2,3}$\footnote{nurlan.qanat@mail.ru},A . A. Pivovarov$^{1}$\footnote{tex$\_$k@mail.ru}\\
\small
\emph{$^{1}$ Joint Intsitute for Nuclear Research, Dubna, 141980, Russia}\\
\small
\emph{$^{2}$ Institute of Nuclear Physics, Almaty, 050032, Kazakhstan}\\
\small
\emph{$^{3}$Eurasian National University, Nur-Sultan, 01008, Kazakhstan}}
\maketitle
\small

\begin{abstract}

In the extended NJL model, the decay widths of $\tau\rightarrow (K, K(1460))\nu_{\tau}$, $(K, K(1460))\rightarrow {\mu}\nu_{\mu}$ are calculated. The contributions from intermediate axial-vector mesons ${K}_{1}(1270)$, ${K}_{1}(1400)$ and the first radially excited state  ${K}_{1}(1650)$ are taken into account. Estimates for the weak decay coefficients ${F}_{K}$ and ${{F}_{{K}^{'}}}$ are given. Predictions are made for the width of $\tau\rightarrow K(1460)\nu_{\tau}$ decay and ${{F}_{{K}^{'}}}$ constant.

\end{abstract}
\large

\section{Introduction}

The decay of ${K}\rightarrow {\mu}\nu_{\mu}$ as well as the weak decay constant ${F}_{K}$ are currently measured experimentally fairly well \cite{Tanabashi:2018oca}. At the same time, the theoretical description of the process $\tau\rightarrow K\nu_{\tau}$ as well as the theoretical calculation of the weak decay constant ${F}_{K}$ are of great interest and are being actively studied recently in various models. For example, in a recent paper \cite{Chang:2018aut} an attempt has been made to describe these quantities by using an improved holographic wave function. In the paper \cite{Choi:2007yu}, a theoretical estimate of the constant ${F}_{K}$ was obtained in the framework of the light-front quark model. A number of attempts were also made to estimate the weak decay constant ${F}_{K}$ by using the lattice calculations \cite{Carrasco:2014poa, Bazavov:2014wgs, Blum:2014tka}.

The constants ${F}_{K}$ and ${F}_{\pi}$ can be calculated on the basis of the study of the $\tau\rightarrow (\pi, K)\nu_{\tau}$ decays. It is especially convenient to do calculation in the framework of the extended NJL model, which allows one to take into account contributions not only from intermediate axial-vector mesons in the ground state but also from their first radial excitations \cite{Volkov:1996br, Volkov:1996fk, Volkov:1997dd, Volkov:2006kw, Volkov:2016umo, Volkov:2017arr}. In this model, the main $\tau$ - lepton decays \cite{Volkov:2017arr, Volkov:2016afr} and some suppressed decays, for example $\tau\rightarrow \pi(\eta, {\eta}^{'})\nu_{\tau}$ \cite{Volkov:2012be}, were described.

In the papers \cite{Volkov:2017arr, Ahmadov:2015zua} the $\tau\rightarrow (\pi, \pi(1300))\nu_{\tau}$ decays were studied and the constants ${F}_{\pi}=94.6$ MeV and ${{F}_{{\pi}^{'}}}=4.7$ MeV were found. The contributions from the intermediate states ${a}_{1}(1260)$ and ${a}_{1}(1640)$ were taken into account. In the present paper, the $\tau\rightarrow (K, K(1460))\nu_{\tau}$ and (${K},{K(1460)})\rightarrow {\mu}\nu_{\mu}$ decays will be described by taking into account the intermediate states ${K}_{1}(1270)$, ${K}_{1}(1400)$ and ${K}_{1}(1650)$.

Unlike the process $\tau\rightarrow (\pi, \pi(1300))\nu_{\tau}$, in the $\tau\rightarrow (K, K(1460))\nu_{\tau}$ decay among the main two intermediate axial-vector physical states ${K}_{1}(1270)$ and ${K}_{1}(1400)$ should be taken into account. This is a consequence of mixing two axial-vector strange mesons ${K}_{1A}$ and ${K}_{1B}$. The first of them has a connection with quarks through the operator ${\gamma}_{5}{\gamma}_{\mu}$, and the second through the operator ${\gamma}_{5} {\partial}_{\mu}$. In the case of the chiral group $ SU (2) \times SU (2) $, such axial-vector mesons ${a}_{1A}$ and ${a}_{1B}$ do not mix with each other. At the same time, for the group $U(3)\times U(3)$ due to a large mass difference between the quarks ${m}_{s} = 420$ MeV, ${m}_{u}={m}_{d}=280$  MeV chiral symmetry is noticeably broken and the states ${K}_{1A}$ and ${K}_{1B}$ begin to mix with each other with a constant connection proportional to the mass difference ${m}_{s}$ , ${m}_{u}={m}_{d}$. This effect will be taken into account in the present work. As a result, the physically observable states are being to have masses $M_{{K}_{1}(1270)}=1272$ MeV and $M_{{K}_{1}(1400)}=1403$ MeV \cite{Tanabashi:2018oca}. These states are related with ${K}_{1A}$, ${K}_{1B}$ as follows (see the papers \cite{Volkov:1984fr, Suzuki:1993yc, Guo:2008sh}):

\begin{displaymath} \label{eq_1}
{K}_{1}(1270)=\sin(\beta){K}_{1A} + \cos(\beta){K}_{1B},
\end{displaymath}
\begin{equation}
{K}_{1}(1400)=\cos(\beta){K}_{1A} - \sin(\beta){K}_{1B}.
\end{equation}

By taking into account mixing of the axial-vector states ${K}_{1A}$ and ${K}_{1B}$, we manage to obtain the value for ${F}_{K}$, which in satisfactory agreement with both experiment and other theoretical works, and also give predictions for the $\tau\rightarrow K(1460)\nu_{\tau}$, ${K(1460)}\rightarrow {\mu}\nu_{\mu}$ decay widths.

\section{ Definition of additional renormalization of the kaon field ($Z_{K}$) in the NJL model }
The Nambu-Jona-Lasinio model established itself in the low-energy meson physics  \cite{Volkov:1984kq, Ebert:1985kz, Volkov:1986zb, Vogl:1991qt, Klevansky:1992qe}. The Lagrangian, describing the interaction of strange mesons with quarks, has the following form \cite{Volkov:1986zb}:

\begin{equation} \label{eq_5}
\Delta L_{int}(q,\bar{q}, K, {K}_{1A}, {K}_{1B}) = \bar{q}\left[
g_{K}i\gamma_{5}\lambda_{K}K
+\frac{g_{K_{1A}}}{2}\gamma^{\mu}\gamma_{5}\lambda_{K}{{K}_{1A}}_{\mu}
\right]q
+\frac{g_{B}}{2}{{K}_{1B}}_{\mu}\bar{q}\left[{\overleftrightarrow{\partial}}_{\mu}\gamma_{5}\lambda_{K}
\right]q.
\end{equation}

Here $q$ and $\bar{q}$ are the quark fields with masses of constituent quarks $m_{u} = m_{d} = 280$ MeV, ${m}_{s} = 420$ MeV. The first two terms of the Lagrangian are related to the standard NJL model, the constants $g_{K} = \left(4I_{2}(m_{u},m_{s})\right)^{-1/2}$, $g_{K_{1A}} = \sqrt{6}g_{K}$. The loop integral $I_{2}(m_{u},m_{s})$ has the form:

\begin{equation}
I_{2}(m_{u}, m_{s})=-i \frac{N_{C}}{{(2\pi)}^{4}} \int \frac{\Theta(\Lambda^2_{4}+k^2)}{(m^2_{u}-k^2)(m^2_{s}-k^2)}d^4k,
\end{equation}
$\Lambda_{4}$ =1250 MeV is the four-dimensional cutoff parameter\cite{Volkov:1986zb}.

The last term of the Lagrangian ($\ref{eq_5}$) goes beyond the standard NJL model and describes the interactions of the meson ${K}_{1B}$ with the quarks including the derivative (${\overleftrightarrow{\partial}}_{\mu}$). Note that in the case of the $SU(2) \times SU(2)$ chiral-symmetric NJL model, there is no mixing between axial-vector mesons ${a}_{1A}$ and ${a}_{1B}$; therefore, when describing $\pi-{a}_{1}$ transitions, the pion has only one partner among the axial-vector mesons, namely ${a} _{1}(1260)$ \cite{Volkov:1986zb}. In the case of kaons, in describing such transitions, an additional partner may appear in the form of the ${K}_{1B}$ meson. The ${K}_{1A}$ and ${K}_{1B}$ mesons begin to mix with each other and as a result, we arrive at two physically observable states ${K}_{1}(1270)$ and ${K}_{1}(1400)$ ($\ref{eq_1}$). The effect of such mixing was first studied within the framework of the NJL model in the paper \cite{Volkov:1984fr} and then re-examined in the papers \cite{Suzuki:1993yc, Guo:2008sh}.

We now find additional renormalization of the kaon field due to a possible kaon transition in the axial-vector state ${K}_{1A}$ and then returning again to the pseudoscalar state $K$. Here we consider possible transformation transitions between pseudoscalar and axial-vector nonnets existing in the standard NJL model, which based on effective four-quark interactions of scalar, pseudoscalar, vector, and axial-vector types without derivatives (${\partial}_{\mu}$). In our case, the state ${K}_{1A}$ is expressed through two physical observable states ${K}_{1}(1270)$ and ${K}_{1}(1400)$. This situation is a consequence of the mixing of the axial-vector mesons ${K}_{1A}$ belonging to the standard NJL model with the axial-vector mesons ${K}_{1B}$, which are related to quarks by ${\gamma}_{5} {\partial}_{\mu}$ and related to the nonet, which is not included in the standard NJL model.

\begin{figure}[H]
\center{\includegraphics[width=0.4\linewidth]{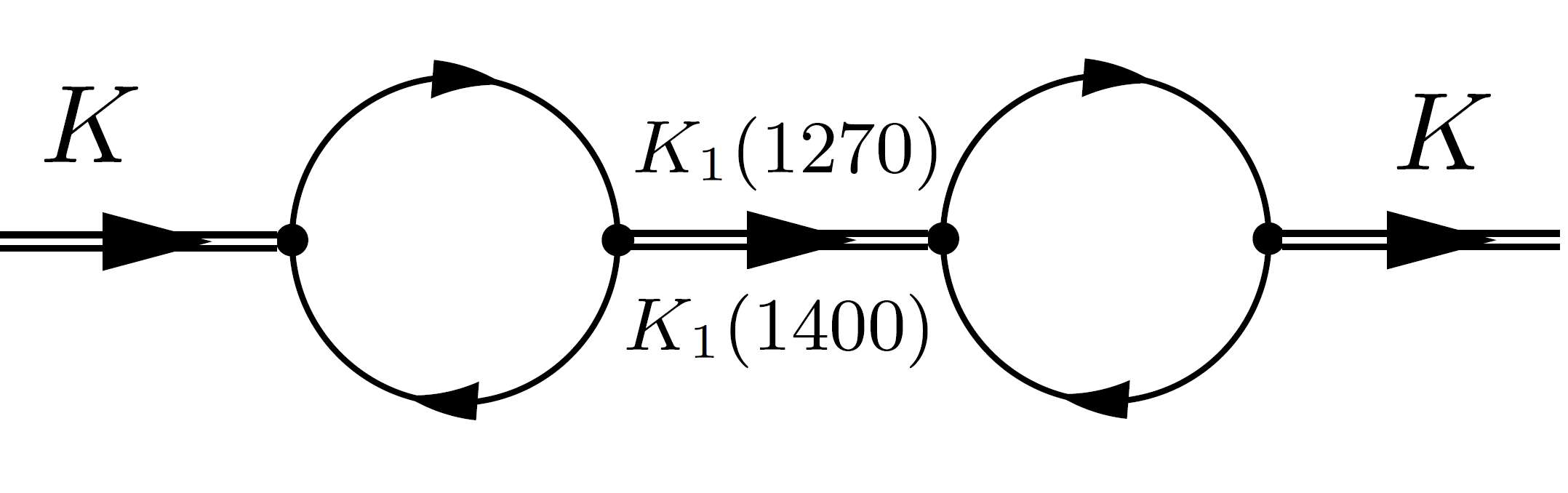}}
\linespread{1} \caption { The $ K-K_{1}-K$ transition's diagram. }
\label{pic_1:pic_1}
\end{figure}

Taking into account the intermediate physical meson states ${K}_{1}(1270)$ and ${K}_{1}(1400)$ will lead to an additional renormalization of the kaon field. As a result, for a constant we get $g_{K} = \left(4I_{2}(m_{u},m_{s})\right / Z_{K})^{-1/2}$, where

\begin{equation}
Z_{K} = \left(1 - \frac{3}{2}(m_{s}+m_{u})^{2}\left(
\frac{{\sin}^{2}(\beta)}{M^{2}_{K_{1}}(1270)}+
\frac{{\cos}^{2}(\beta)}{M^{2}_{K_{1}}(1400)}
\right)\right)^{-1}=
\left(1 - \frac{3}{2}
\frac{(m_{s}+m_{u})^{2}}{M^{2}_{K_{1A}}}
\right)^{-1}
\approx 1.76.
\end{equation}

Here, for convenience in our further calculations, we have introduced the effective mass for the intermediate nonphysical meson ${K}_{1A}$:

\begin{displaymath}
M_{K_{1A}} = \left(\frac{\sin^{2}(\beta)}{M^{2}_{K_{1}(1270)}} + \frac{\cos^{2}(\beta)}{M^{2}_{K_{1}(1400)}}\right)^{-1/2} \approx \: 1305 MeV,
\end{displaymath}
here the angle is $\beta=57^{\circ}$ \cite{Volkov:1984fr, Suzuki:1993yc}.

It is interesting to note that with such a value of $Z_{K}$ and the mass of the constituent quark ${m}_{s} = 420$ MeV one obtains good agreement with the experimental value for the kaon mass. Indeed, using the Gell-Mann-Oakes-Renner formula for the kaon mass, we obtain:

\begin{equation}
M^{2}_{K}=\frac{{g_{K}}^{2}}{G_{1}}\left[ 1-4G_{1}(I_{1}(m_{s}) + I_{1}(m_{u})) \right] + Z_{K}(m_{s}-m_{u})^{2},
\end{equation}

\begin{displaymath}
M_{K}=494.8 \: MeV,
\end{displaymath}
where $G_{1}=4.9$ ${GeV}^{-2}$ is the interaction constant of two pseudoscalar currents in the initail effective 4-quark lagrangian of the NJL model \cite{Volkov:1986zb}. The integrals $I_{1}(m_{s}), I_{1}(m_{u})$ are square-divergent:

\begin{equation}
I_{1}(m_{q})=-i \frac{N_{C}}{{(2\pi)}^{4}} \int \frac{\Theta(\Lambda^2_{4}+k^2)}{(m^2_{q}-k^2)}d^4k
\end{equation}

The experimental mass value \cite{Tanabashi:2018oca}:

\begin{displaymath}
{M_{K}}^{exp}=493.67 MeV .
\end{displaymath}

\section{ Decay width of $\tau\rightarrow K \nu_{\tau}$ in the standard NJL model }

To describe the decay $\tau\rightarrow K \nu_{\tau}$ in addition to the Lagrangian of the standard NJL model with the quark-kaon interaction constant $g_{K}$, here we also need the Lagrangian for the weak interaction of the lepton current with quarks:

\begin{equation}
 L^{weak}=\bar{\tau}\gamma_{\mu}(1-\gamma_{5})\nu \frac{G_{F}}{\sqrt{2}}|V_{us}|\bar{s}(1-\gamma_{5})\gamma_{\mu}u,
\end{equation}
where $G_{F}$ is the Fermi constant, $|{V}_{us}|$ is the element of the Cabibbo-Kobayashi-Maskawa matrix.

The diagrams describing this decay are shown in Figures 2 and 3.

\begin{figure}[h]
\center{\includegraphics[width=0.4\linewidth]{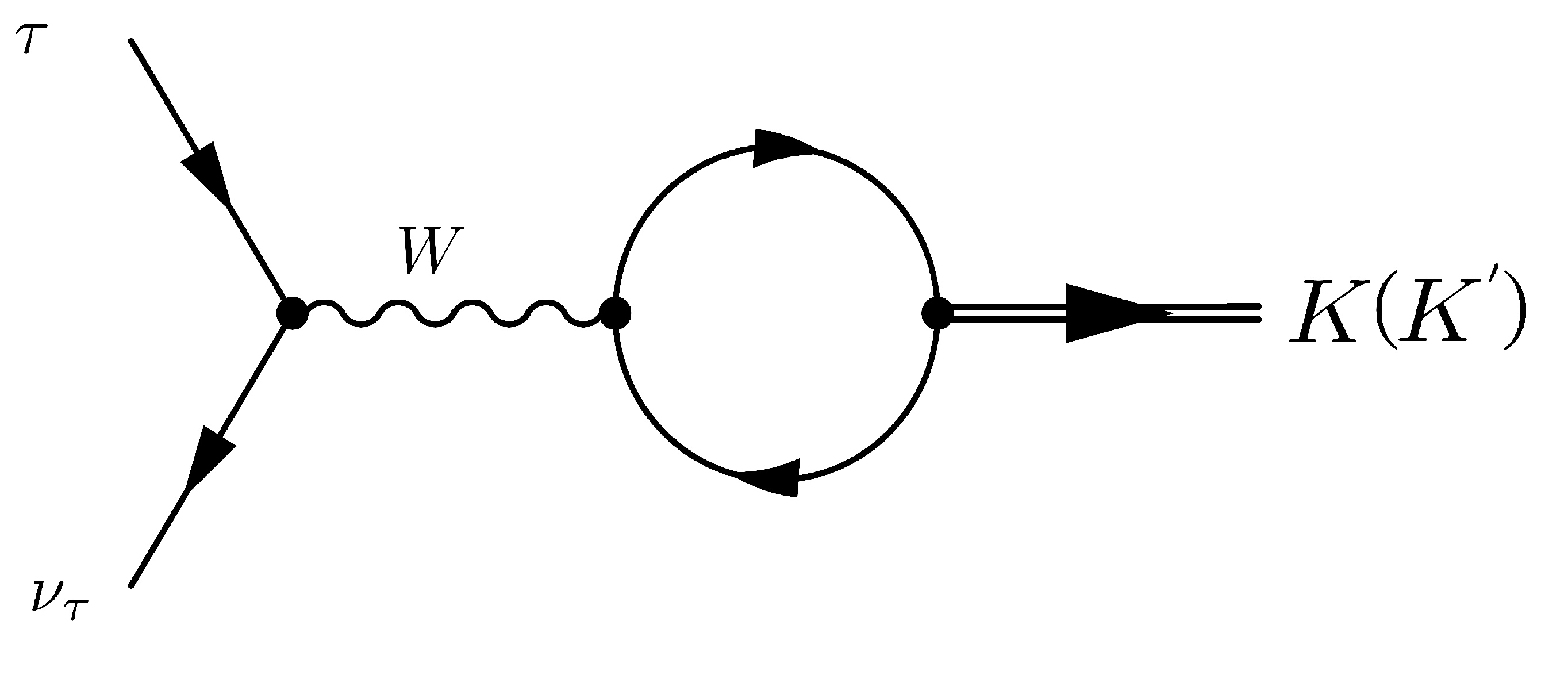}}
\linespread{1} \caption {Contact diagram of the decay $\tau\rightarrow K \nu_{\tau}$. }
\label{pic_2:pic_2}
\end{figure}

\begin{figure}[h]
\center{\includegraphics[width=0.5\linewidth]{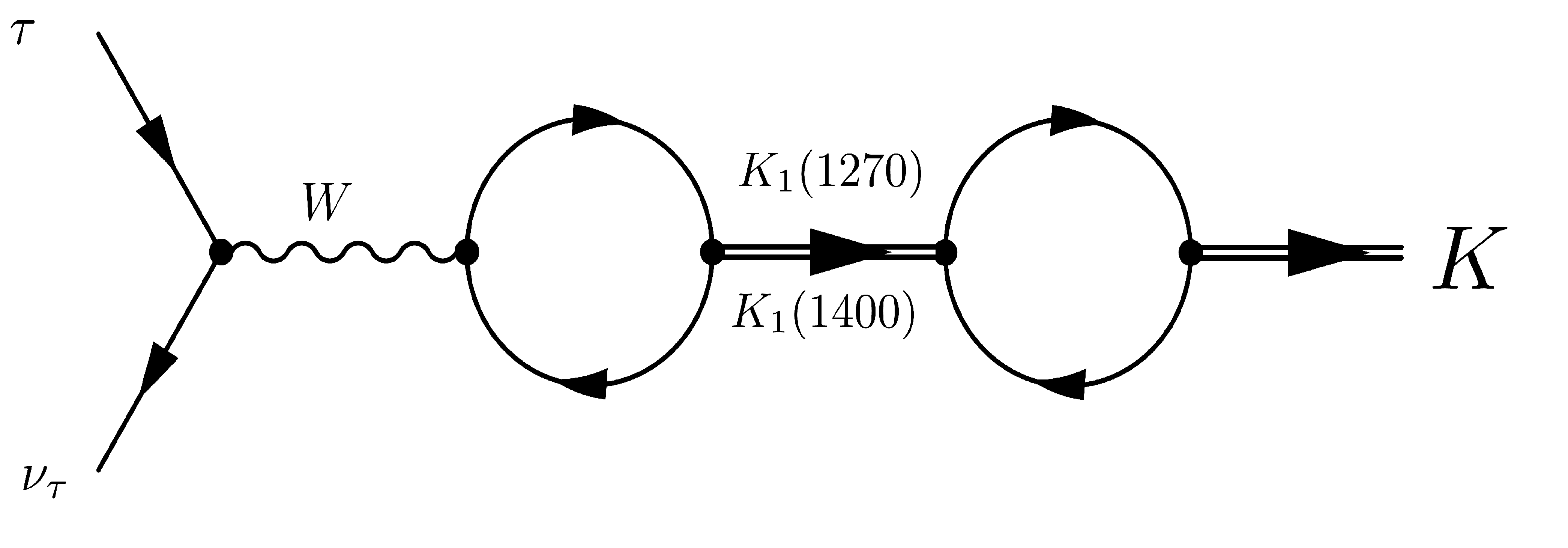}}
\linespread{1} \caption {Diagram with the intermediate axial vector mesons ${K}_{1}(1270)$ and ${K}_{1}(1400)$.}
\label{pic_3:pic_3}
\end{figure}

The corresponding amplitude in the standard NJL model takes the form:

\begin{equation}
A_{\mu} = \frac{G_{F}}{\sqrt{2}}|V_{us}|L_{\mu}\left( A_{C} + A_{K_{1A}} \right)p_{\mu}.
\end{equation}

Separate contributions are:

\begin{displaymath}
A_{C} = \sqrt{2}Z_{K}\frac{m_{s}+m_{u}}{2g_{K}},
\end{displaymath}

\begin{displaymath}
A_{K_{1A}} = \sqrt{2}Z_{K}\frac{m_{s}+m_{u}}{2g_{K}}\left( -\frac{3{(m_{s}+m_{u})}^{2}}{M^{2}_{K_{1A}}} \right)=
\sqrt{2}\frac{m_{s}+m_{u}}{2g_{K}}(1-Z_{K}).
\end{displaymath}

The constant $Z_{K}$ appearing in the numerator of the contact amplitude is completely reduced with the corresponding constant appearing in the amplitude with the intermediate $K_{1}$ meson. A similar situation occurred in the $\tau\rightarrow \pi \nu_{\tau}$ \cite{Volkov:2017arr, Volkov:1986zb}. Representing the hadron current in the form $\sqrt{2}F_{K}p_{\mu}$, for the weak decay constant $F_{K}$ we get the following expression:
\begin{equation}
F_{K} = \frac{m_{s}+m_{u}}{2g_{K}}.
\end{equation}

As a result for the decay width and the constant ${F}_{K}$ we get the values:
\begin{displaymath}
\Gamma_{NJL}(\tau\rightarrow K \nu_{\tau})=1.19\times10^{-11} MeV,
\end{displaymath}

\begin{displaymath}
F_{K}=95 \: MeV.
\end{displaymath}

Experimental values are:

\begin{equation} \label{eq_2}
\Gamma_{exp}(\tau\rightarrow K \nu_{\tau})=(1.58 \pm 0.02)\times10^{-11} MeV \: \cite{Tanabashi:2018oca},
\end{equation}
\begin{equation} \label{eq_3}
\Gamma_{exp}(\tau\rightarrow K \nu_{\tau})=(1.59 \pm 0.06)\times10^{-11} MeV \: \cite{Schael:2005am},
\end{equation}

\begin{displaymath}
{F_{K}}_{exp}=110.02 MeV \: \cite{Tanabashi:2018oca}.
\end{displaymath}

\section{The decays $\tau\rightarrow (K , K(1460))\nu_{\tau}$ in the extended NJL model  }

In the extended NJL model, the Laganian for the mesons $K$,  $K(1460)$, $K_{1A}$, $K_{1}(1650)$ has the form \cite{Volkov:2017arr}:

\begin{displaymath}
\begin{gathered}
\Delta L_{int}(q,\bar{q},K , K(1460), K_{1A}, K_{1}(1650)) = \bar{q} \left[
B_{K}i\gamma_{5}\lambda_{K}K+
B_{K^{'}}i\gamma_{5}\lambda_{K}K^{'}+\right. \\
\left. +\frac{B_{K_1}}{2}\gamma^{\mu}\gamma_{5}\lambda_{K}{K_{1A}}_{\mu}+
\frac{B_{K^{'}_{1}}}{2}\gamma^{\mu}\gamma_{5}\lambda_{K}{K^{'}_{1A}}_{\mu} \right] q,
\end{gathered}
\end{displaymath}
the excited states of the mesons are indicated by a stroke,

\begin{displaymath}
B_{M} = \frac{1}{\sin(2\theta_{M}^{0})}\left[g_{M}\sin(\theta_{M} + \theta_{M}^{0}) +
g_{{M}^{'}}f_{M}({k_{\perp}}^{2})\sin(\theta_{M} - \theta_{M}^{0})\right],
\end{displaymath}

\begin{displaymath}
B_{{M}^{'}} = \frac{-1}{\sin(2\theta_{M}^{0})}\left[g_{M}\cos(\theta_{M} + \theta_{M}^{0}) +
g_{{M}^{'}}f_{M}({k_{\perp}}^{2})\cos(\theta_{M} - \theta_{M}^{0})\right],
\end{displaymath}
$f\left({k_{\perp}}^{2}\right) = 1 + d_{us} {k_{\perp}}^{2}$ is the form factor describing the first radially excited states. Transverse relative momentum of quarks in the meson:

\begin{displaymath}
k_{\perp}=k-\frac{(kp)p}{p^2},
\end{displaymath}
where $p$ is meson momentum, $d_{us}$ is the slope parameter. The slope parameter $d_{us}$ is chosen so that the radially-excited state does not influenced quark condensate and, hence, the values of the constituent quark masses, $\theta_{K}$ and $\theta_{K_{1A}}$ are the mixing angles of the mesons in the ground and excited states. The determination of the mixing angles can be found in \cite{Volkov:2017arr} (Chapter 2.2),

\begin{displaymath}
d_{us} = -1.762 \textrm{GeV}^{-2},
\end{displaymath}

\begin{displaymath}
\theta_{K} = 58.11^{\circ}, \: \theta_{K_{1A}} = 85.97^{\circ},
\end{displaymath}

The auxiliary quantities $\theta_{M}^{0} $ are expressed in terms of integrals:

\begin{displaymath}
\sin(\theta_{M}^{0})=\sqrt{\frac{1+R_{M}}{2}},
\end{displaymath}
where

\begin{displaymath}
R_{K}=\frac{{I_{2}}^{f}}{\sqrt{Z_{K}I_{2}{I_{2}}^{f^{2}}}},
R_{K_{1A}}=\frac{{I_{2}}^{f}}{\sqrt{I_{2}{I_{2}}^{f^{2}}}}.
\end{displaymath}

\begin{figure}[H]
\center{\includegraphics[width=0.5\linewidth]{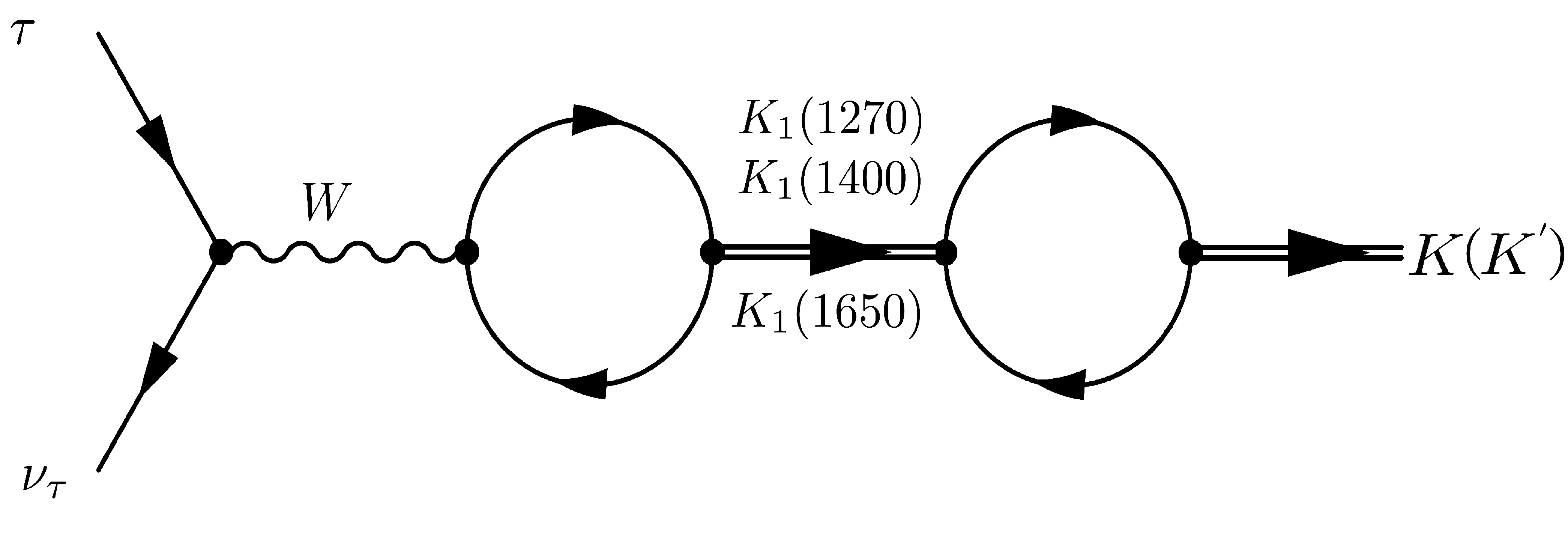}}
\linespread{1} \caption { Diagram with intermediate axial-vector mesons.}
\label{pic_4:pic_4}
\end{figure}

The quark-meson interaction constants for excited mesons:
\begin{displaymath}
{g}_{K^{'}} = \left(4{I_{2}}^{f^{2}_{us}}(m_{u},m_{s})\right)^{-1/2},
\quad g_{K^{'}_{1A}} = \left(\frac{2}{3}I_{2}^{f_{us}^{2}}(m_{u},m_{s})\right)^{-1/2}.
\end{displaymath}
where the integral $I_{2}^{f_{us}^{n}}(m_{u},m_{s})$ has the following form:

\begin{displaymath}
I_{2}^{f_{us}^{n}}(m_{u},m_{s})=-i \frac{N_{C}}{{(2\pi)}^{4}} \int \frac{f_{us}^{n}(k^2_{\perp})}{(m^2_{u}-k^2)(m^2_{s}-k^2)}\Theta(\Lambda^2_{3}-k^2_{\perp})d^4k,
\end{displaymath}
$\Lambda_{3}$ =1030 MeV is the three-dimensional cutoff parameter. All parameters of the extended NJL model are  earlier defined and taken from the paper \cite{Volkov:2017arr}.

The diagrams, describing the $\tau\rightarrow (K , K(1460))\nu_{\tau}$ decays, are shown in Figures 2 and 4. The corresponding amplitude in the extended NJL model takes the form:

\begin{equation}
A_{\mu} = G_{F}V_{us}L_{\mu}\left( A_{C} + A_{K_{1A}} + e^{i\phi}A_{K^{'}_{1A}} \right)p_{\mu}.
\end{equation}

The extended NIL model does not pretend to a correct description of the relative phases between the ground and excited states of intermediate mesons. Therefore, we selected them based on experimental data, as was done on our previous works \cite{Volkov:2017arr}.

The contribution from the contact diagram (Fig.2) is:
\begin{displaymath}
 A_{C} = Z_{K}\frac{m_{u} + m_{s}}{2g_{K}}C_{K},
 \end{displaymath}

The contributions from the diagrams with intermediate axial-vector mesons (Fig.4) are:

\begin{displaymath}
 A_{K_{1A}} = - \frac{6{(m_{s} + m_{u})}^{3}}{2M^{2}_{K_{1A}}} \frac{C_{K_{1A}}}{g_{K_{1A}}} {I_{2}}^{K_{1A}K},
 \end{displaymath}

\begin{displaymath}
 A_{K^{'}_{1A}} = - \frac{6{(m_{s} + m_{u})}^{3}}{2M^{2}_{K_{1}(1650)}} \frac{C_{K^{'}_{1A}}}{g_{K_{1A}}} {I_{2}}^{K^{'}_{1A}K},
 \end{displaymath}
here $M_{K_{1}(1650)}=1672$ MeV is the mass of a radially excited meson \cite{Tanabashi:2018oca}. The coefficients describing $W$ - boson transitions to mesons are:

\begin{displaymath}
C_{M} = \frac{1}{\sin\left(2\theta_{M}^{0}\right)}\left[\sin\left(\theta_{M} + \theta_{M}^{0}\right) +
R_{M}\sin\left(\theta_{M} - \theta_{M}^{0}\right)\right],
\end{displaymath}

\begin{displaymath}
C_{M^{'}} = \frac{-1}{\sin\left(2\theta_{M}^{0}\right)}\left[\cos\left(\theta_{M} + \theta_{M}^{0}\right) +
R_{M}\cos\left(\theta_{M} - \theta_{M}^{0}\right)\right].
\end{displaymath}

Using the resulting hadron current, similar calculations can be performed for the decays $\tau\rightarrow K^{'}\nu_{\tau}$, ${K}\rightarrow {\mu}\nu_{\mu}$, ${K^{'}}\rightarrow {\mu}\nu_{\mu}$. In this case, we consider three cases of choosing the relative phase of the intermediate axial-vector meson $K_{1}(1650)$. The theoretical estimates obtained for the decay widths and the weak kaon decay constant are shown in Table 1. The value $\phi=102^{\circ}$ gives us the best agreement with the experimental data.

\begin{table}[H]
\begin{center}
\begin{tabular}{|c|c|c|c|}
\hline
& \multicolumn{3}{c|}{NJL-predicted width, MeV} \\
\cline{2-4}
\raisebox{1.5ex}[0cm][0cm]{Decay}
& $\phi=0^{\circ}$& $\phi=180^{\circ}$& $\phi=102^{\circ}$ \\
\hline
$\tau\rightarrow K\nu_{\tau}$ & $1.306\times 10^{-11}$ & $1.77\times 10^{-11}$ & $1.598\times 10^{-11}$ \\
$\tau\rightarrow K^{'}\nu_{\tau}$ & $2.09\times 10^{-14}$ & $2.164\times 10^{-13}$ & $1.426\times 10^{-13}$ \\
$K\rightarrow \mu\nu_{\mu}$ & $2.761\times 10^{-14}$ & $3.758\times 10^{-14}$& $3.379\times 10^{-14}$ \\
$K^{'}\rightarrow \mu\nu_{\mu}$ & $1.145\times 10^{-15}$ & $1.186\times 10^{-14}$ & $7.811\times 10^{-15}$ \\
$F_{K}$ & $100$ & $116$ & $110.14$ \\
$F_{K^{'}}$ & $11.3$ & $36.3$ & $29.54$ \\
\hline
\end{tabular}
\end{center}
\linespread{1} \caption {
There are the obtained decay widths and weak decay constants of the kaon in the framework of the extended NJL model. The values ​​of case I correspond to the choice of the phase $\phi=0^{\circ}$ of the radially excited meson ${K}_{1}(1650)$. Accordingly, with the phases $\phi=180^{\circ}$ and $\phi=102^{\circ}$, we obtain the results of case II and case III.}
\end{table}

\begin{displaymath}
\Gamma_{exp}(K\rightarrow \mu\nu_{\mu})=(3.38 \pm 0.006)\times 10^{-14} MeV\cite{Tanabashi:2018oca}.
\end{displaymath}

\section{Conclusion and Discussion}
In this work, the decay $\tau\rightarrow K\nu_{\tau}$ has been described within the framework of the standard and extended NJL models. An additional renormalization constant $Z_{K}$ was determined, which arises when taking into account a possible $K_{1} -K$ transition in a kaon. These transitions are considered only inside the pseudoscalar and axial-vector nonets, which are in the standard NJL model. In determining the renormalization constant $Z_{K}$, the masses of the physical observable states of the axial vector mesons $K_{1}(1270)$ and $K_{1}(1400)$, which are the product of the mixing of the nonphysical states $K_{1A}$ and $K_{1B}$, were taken into account.

The individual contributions to the amplitude and width of the $\tau\rightarrow K\nu_{\tau}$ decay from both the contact term and the channels with intermediate axial-vector mesons $K_{1}(1270)$, $K_{1}(1400)$ and $K_{1}(1650)$ were estimated. The dominant contribution to the determination of the $F_{K}$ is the contribution from the contact diagram, where as, the contributions from intermediate axial-vector mesons are smaller. Our calculations show that when describing the $\tau\rightarrow K^{'}\nu_{\tau}$ decay, the role of the contributions of the contact and axial-vector channels changes and the channel with the $K_{1}(1270)$ meson is decisive. The best agreement with the experimental data was obtained in the extended NJL model with the phase $\phi=102^{\circ}$ of the intermediate axial-vector meson $K_{1}(1650)$.

The accuracy of the calculation in the NJL model is determined by the accuracy of the conservation of chiral symmetries on which the NJL model is based. This accuracy depends on the preservation of the axial current (PCAC principle). In the case of $ SU(2) \times SU(2)$ symmetries, this accuracy is determined by the formula: $ M^2_{\pi}/M^2_{N} \approx 0.02$ \cite{Vainshtein:1970zm}. In the case of $ U(3) \times U(3)$ chiral symmetry, this violation turns out to be stronger since $ M^2_{K}/M^2_{\Sigma} \approx 0.17$. Our previous results show that for processes involving strange particles it is possible to obtain results within the specified accuracy \cite{Volkov:2006kw, Volkov:2016umo, Volkov:2017arr, Volkov:1993jw, Ebert:1994mf}.

Recently, similar descriptions of $F_{K}$ have been studied in other theoretical models. For example, in the work \cite{Chang:2018aut} similar deductions were made within the model using the holographic wave function $F_{K} = 108.47 $ MeV and $F_{K} = 110.59 $ MeV. In the paper \cite{Choi:2007yu}, anologous calculations were given in the quark model on the light cone and the estimates $ F_{K} = 113.84 $ MeV and $ F_{K} = 109.6 $ MeV were obtained. Also, calculations performed on the lattice gave the result $ F_{K} = 109.17 $ MeV \cite{Carrasco:2014poa}, $ F_{K} = 110.25 $ MeV \cite{Bazavov:2014wgs}, $ F_{K} = 109.96 $ MeV \cite{Blum:2014tka}.

\section*{Acknowledgments}

We are greatful to A.B. Arbuzov and A.A. Osipov for useful disscussions; this work is supported by the JINR grant for young scientists and specialists No. 19-302-06.

\end{document}